\documentclass[twocolumn,prc,preprintnumbers,nofootinbib,showpacs]{revtex4}
\usepackage{graphicx}
\usepackage{bm}
\usepackage{amssymb,amsmath}

\usepackage{array}

\def\beqn{\begin{eqnarray}}
\def\eeqn{\end{eqnarray}}
\def\barr{\begin{array}}
\def\earr{\end{array}}
\def\btab{\begin{tabular}}
\def\etab{\end{tabular}}
\def\bite{\begin{itemize}}
\def\eite{\end{itemize}}
\def\bcen{\begin{center}}
\def\ecen{\end{center}}

\def\eq{\begin{equation}}
\def\ee{\end{equation}}
\def\nn{\nonumber}

\def\pgdagger{P\hspace{-0.27cm}/}

\def\keldagger{k\hspace{-0.2cm}/}

\def\q2dagger{q_2\hspace{-0.35cm}/\;}

\begin{document}


\title{Analyzing power in elastic scattering of the electrons off a spin-0 target}

\author{Mikhail Gorchtein} 
\email{gorshtey@caltech.edu}
\author{Charles J. Horowitz} 
\email{horowit@indiana.edu}
\affiliation{Nuclear Theory Center and Department of Physics, 
Indiana University, Bloomington, IN 47408, USA} 

\date{\today}

\begin{abstract}
We consider the analyzing power on a spin-0 nuclear target. This observable 
is related to the imaginary part of the two-photon-exchange (box) diagram. 
We consider the contributions of elastic and inelastic intermediate states. 
The former requires knowledge of the elastic nuclear form factor, while the 
latter uses the optical theorem as input. The elastic contribution scales as 
the nuclear charge $Z$, while the inelastic contribution as the ratio of 
the atomic number and nuclear charge, $A/Z$. We provide estimates for 
$^4$He and $^{208}$Pb, in the kinematics of existing or upcoming experiments. 
In both cases, we predict negative values for the analyzing power 
of a few parts per million, and 
the dominant contribution is due to inelastic intermediate states. 
The analyzing power can contribute a substantial systematic error in 
parity-violating experiments.
\end{abstract}


\maketitle

\section{Introduction}

Over the past few years, much attention was paid to the two-photon-exchange 
(TPE) effects in elastic electron scattering off nucleons and nuclei. The 
discrepancy between the values 
of the elastic form factor ratio of the proton, $G_E/G_M$ obtained with the 
Rosenbluth separation technique \cite{gegm_ros} on one hand, 
and the polarization transfer technique \cite{gegm_pol} on the other hand, 
is believed to be due to these TPE effects \cite{marcguichon}. To ultimately 
disentangle these effects, two experiments are planned at JLab \cite{epm_jlab} 
and at VEPP ring \cite{epm_vepp} that will measure the ratio of the electron 
and positron cross sections.

Another way to measure the TPE effects is to study the analyzing power, called
Mott asymmetry in low energy polarimetry. This asymmetry involves a 
transversely polarized beam of electrons. Because of time reversal symmetry, 
a non-zero 
asymmetry requires a non-zero imaginary part of the elastic amplitude and
is due to exchange of at least two photons. This observable scales naively 
as $\frac{m_e}{E} Z\alpha_{em}$, with $m_e$ the electron mass, $E$ the beam 
energy, $Z$ the charge of the target particle, and 
$\alpha_{em}$ the fine structure constant. A rough estimate gives $10 ppm$ 
for the case of $500$ MeV beam scattering off the proton target.

Parity violating experiments use a longitudinally polarized beam of electrons 
and measure the difference in cross section due to flipping the beam 
polarization. Such PV asymmetries are typically of order $1 ppm$. 
It can be seen, that a small 
transverse component of the electron spin can lead to a substantial 
systematical effect on the PV asymmetry. On the other hand, the analyzing power 
can be measured easily with the same apparatus used in PV experiments. 
There exist several measurements of this effect \cite{bn_exp} and a number 
of theoretical estimates \cite{derujula}, \cite{bn_theo} for the proton target. 
In the case of a nuclear target, the effects of the exchange of two photons is 
expected to be even more important, as it grows with the nuclear charge $Z$. 
This paper is dedicated to calculating the analyzing power on two spin-0 
nuclear targets used in two PV experiments runnning at JLab.
HAPPEX experiment \cite{HAPPEX} uses the $^4$He target and 3 GeV electron beam. 
PREX experiment \cite{PREX} uses the $^{208}$Pb target and 850 MeV electrons.
The paper is organized as follows. We start with defining the kinematics and 
conventions in Section \ref{kinematics}. In Section \ref{im}, we calculate the 
imaginary part of the elastic electron-nucleus amplitude due to elastic and 
inelastic intermediate states. In Section \ref{results}, we present the results 
of our calculation for $^4$He and $^{208}$Pb, and discuss their implications 
for the experiment.

\section{Kinematics and observables}
\label{kinematics}
Kinematics of elastic electron-nucleus scattering process 
$e(k)+N(p)\to e(k')+N(p')$ is fixed by three independent vectors,
\beqn
P&=&\frac{p+p'}{2}\nn\\
K&=&\frac{k+k'}{2}\nn\\
q&=&k-k'\;=\;p'-p,
\eeqn
and two independent Mandelstam invariants $Q^2=-q^2>0$ and $\nu=(P\cdot K)/M$, 
where $M$ denotes the mass of the nucleus. The usual polarization 
parameter $\varepsilon$ of the virtual photon can be related to the 
invariants $\nu$ and $Q^2$ (neglecting the mass of the electron here):
\beqn
\varepsilon\,=\,\frac{\nu^2-M^2\tau(1+\tau)}{\nu^2+M^2\tau(1+\tau)},
\eeqn
with $\tau=Q^2/(4M^2)$. Elastic scattering of electrons off a spin-less nuclei
is described by two amplitudes,
\beqn
T&=&\frac{e^2}{Q^2}
\bar{u}(k')
\left\{ m_e A_1\,+\,A_2\pgdagger\,\right\}
u(k)
\label{eq:ampl}
\eeqn
\indent
The amplitudes $A_{1,2}$ are functions of the invariants $\nu,Q^2$.
In the one-photon exchange (OPE) approximation, the helicity-flip amplitude 
$A_1$ vanishes, while the amplitude $A_2$ is related to the elastic nuclear 
form factor that only depends on $t$:
\beqn
A_2^{(0)}\;=\;2ZF_N(Q^2)
\eeqn
\noindent
with $Z$ the nuclear charge. The unpolarized cross section is given by
\beqn
\frac{d\sigma}{d\Omega_{Lab}}&=&F_N^2(Q^2)
\frac{d\sigma_0}{d\Omega_{Lab}},
\eeqn
with the usual Rutherford cross section 
\beqn
\frac{d\sigma_0}{d\Omega_{Lab}}&=&\frac{4\alpha^2Z^2\cos^2\frac{\Theta}{2}}{Q^4}
\frac{E'^3}{E},
\eeqn
$\Theta$ the electron Lab scattering angle and $E(E')$ the incoming (outgoing) 
electron Lab energy. 
The analyzing power, or beam normal spin asymmetry is defined as
\beqn
A_n\;=\;
\frac{\sigma_\uparrow-\sigma_\downarrow}{\sigma_\uparrow+\sigma_\downarrow}
\,,\eeqn
\noindent
where $\sigma_\uparrow$ ($\sigma_\downarrow$) denotes the respective elastic 
cross 
section with the incoming electrons polarized along the positive (negative) 
normal vector $S^\gamma$, 
\beqn
S^\gamma\;=\;\varepsilon_{\alpha\beta\gamma\delta}P^\alpha K^\beta q^\delta
\eeqn

This observable requires a non-zero imaginary part of the elastic amplitude, 
thus it is identically zero in the OPE approximation. Including the exchange 
of two photons, we obtain to leading order in $\alpha_{em}$
\beqn
A_n\;=\;-\frac{m_e}{\sqrt{s}}\tan\left(\frac{\theta_{cm}}{2}\right)
\frac{{\rm Im}A_1}{ZF_N(Q^2)}\,,\label{eq:an_general}
\eeqn
with ${\rm Im}A_1\sim O(\alpha_{em})$.

\section{Imaginary part of the TPE ampltude}
\label{im}

The imaginary part of TPE amplitude is given by 
\beqn
{\rm Im}T_{2\gamma}
\,=\,e^4\frac{1}{(2\pi)^3}\int\frac{d^3\vec{k}_1}{2E_1}
\frac{1}{Q_1^2Q_2^2}l_{\mu\nu}\cdot W^{\mu\nu},
\label{eq:impart}
\eeqn
where we explicitly set the intermediate electron on-shell, 
$E_1=\sqrt{\vec{k}_1^2+m_e^2}$. 
The leptonic tensor is given by
\beqn
l_{\mu\nu}=\bar{u}(k')\gamma_\nu(\keldagger_1+m_e)
\gamma_\mu u(k).
\eeqn

\subsection{Elastic contribution}
In the case of the elastic 
intermediate state (cf. Fig. 1), the hadronic tensor is 
\beqn
W^{\mu\nu}&=&\pi\delta((P+K-k_1)^2-M^2)\\
&\times&(2p+q_1)^\mu (2p'+q_2)^\nu Z^2F_N(Q_1^2)F_N(Q_2^2)\nn
\eeqn

Above, $q_1^\mu=k-k_1$ denote the incoming and $q_2^\mu=k'-k_1$ the 
outgoing photon momenta, and $Q_{1,2}^2=-q_{1,2}^2$, respectively.
Gauge invariance of the leptonic tensor leads to 
$q_1^\mu l_{\mu\nu}=q_2^\nu l_{\mu\nu}=0$.
For the imaginary part, 
the form factors $F_N$ are the on-shell form factors, and 
we will use experimental fits for them.
\begin{figure}[h]
{\includegraphics[height=1.5cm]{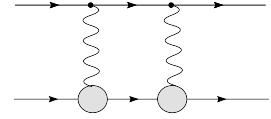}}
\caption{The nucleus box-graph. The shaded blobs represent the nuclear form 
factor}
\label{fig:2gammadiag}
\end{figure}
Evaluating the remaining $\delta$-function in the c.m. frame, we are left 
with the integral over electorn's solid angle $\Omega_1$,
\beqn
{\rm Im}T^{el}_{2\gamma}
=\frac{Z^2e^4}{8\pi^2}\frac{E_1}{\sqrt{s}}\int\!\!
\frac{d\Omega_1}{Q_1^2Q_2^2}l_{\mu\nu}p^\mu p'^\nu F_N(Q_1^2)F_N(Q_2^2),
\label{eq:impart_el}
\eeqn
with the invariant $s=(P+K)^2=M^2+2M\nu+Q^2/2$, and 
$E_1=\frac{s-w^2}{2\sqrt{s}}$ denoting the c.m. energy of the intermediate 
electron. $w^2$ stands for 
the invariant mass squared of the intermediate hadronic 
state. It equals to $M^2$ for the elastic, and lies between the threshold 
for pion production $(M+m_\pi)^2$ and the full energy $s$ for inelastic 
intermediate states.

The integral over the intermediate electron's solid angles can be rewritten 
in terms of the exchanged photons' virtualities $Q_{1,2}^2$:
\beqn
\int d\Omega_1\;=\;\frac{1}{EE_1}\int_0^{4EE_1}\!\!\!\!\!\!dQ_1^2
\int_{Q_-}^{Q_+}\!\!\!\!\frac{dQ_2^2}{\sqrt{(Q_+-Q_2^2)(Q_2^2-Q_-)}}
\eeqn

The limits of the integration $Q_{\pm}$ are given by 
\beqn
Q_\pm&=&\frac{E_1}{E}Q^2+Q_1^2-\frac{Q^2Q_1^2}{2E^2}\\
&\pm& 2\sqrt{Q^2Q_1^2}
\sqrt{\frac{E_1}{E}\left(1-\frac{Q^2}{4E^2}\right)
\left(1-\frac{Q_1^2}{4EE_1}\right)}\nn
\eeqn

Fig. 2 displays the area of the accessible values of $Q_{1,2}^2$ 
for different kinematics and for the case of the nucleon target. The upper 
panels display the case of the elastic intermediate state, while the lower 
panel show the inelastic case for two specific values of $w^2$. 
\begin{figure}[h]
{\includegraphics[height=8.5cm]{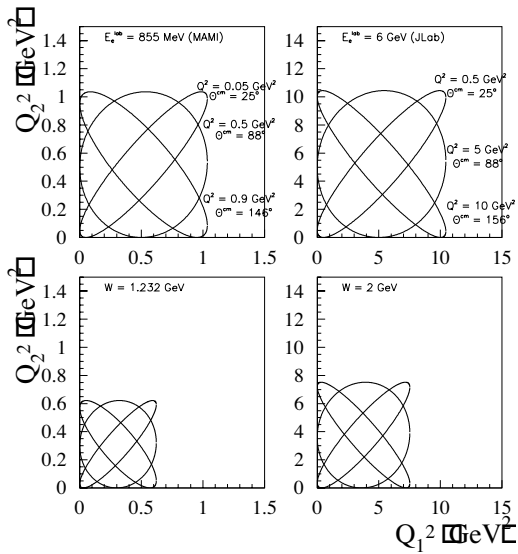}}
\caption{Allowed values of the exchanged photon virtualities $Q_{1,2}^2$ are 
restricted to be inside the ellipses.}
\label{fig:kinbounds}
\end{figure}
By means of standard methods including Dirac algebra and the reduction of 
vector 4-point integrals to scalar 3- and 4-point integrals, we can identify 
the imaginary part of the amplitude $A_1$
\beqn
{\rm Im}A_1^{el}&=&\frac{Z^2\alpha}{\pi}\frac{Q^2}{Q^2-\frac{(s-M^2)^2}{s}}
\frac{s+M^2}{s-M^2}
\label{eq:ima1_elast}\\
&\times&
\int_0^{4E^2}\!\!\!\!\!\!dQ_1^2
\int_{Q_-}^{Q_+}\!\!\!\!\frac{dQ_2^2}{\sqrt{(Q_+-Q_2^2)(Q_2^2-Q_-)}}\nn\\
&\times&
\frac{Q^2-Q_1^2-Q_2^2}{2Q_1^2Q_2^2}F_N(Q_1^2)F_N(Q_2^2)\nn
\eeqn

We notice that the analyzing power does not contain any IR divergencies, so 
that the integrand in the above formula is regular for any allowed values of 
$Q_{1,2}^2$.

\subsection{Inelastic contribution}

We will next estimate the contribution of the inelastic intermediate states 
to the imaginary part of $A_1$ in the case of forward scattering angles. 
We can provide a realistic estimate for the case of nearly forward scattering, 
as it was proposed for the proton target in Refs. \cite{afanasev1}, 
\cite{javvcs}. 
In the forward direction, the imaginary part of the 
doubly virtual Compton scattering amplitude is given in terms of the 
structure functions $W_{1,2}$, and making use of Callan-Gross relation, we have
\beqn
W^{\mu\nu}&=&\pi W_1(w^2,Q_1^2)\\
&\times&
\left\{-g^{\mu\nu} + \frac{P^\mu q_1^\nu + P^\nu q_2^\mu}{(P\tilde{K})}
- \frac{(q_1q_2)}{(P\tilde{K})^2}P^\mu P^\nu\right\}\nn
\eeqn

The structure function $W_1$ is related to the virtual photon cross section,
\beqn
W_1\,=\,\frac{w^2-M^2}{2\pi e^2}\sigma_{\gamma^*N}(w^2,Q_1^2)
\eeqn

In Ref. \cite{javvcs}, it was shown that for the analyzing power 
at very forward angles, the 
$Q_{1,2}^2$ dependence of the cross section can be neglected, as it leads to 
corrections in powers of $Q^2/k^2$, thus 
\beqn
W_1\,\approx\,\frac{w^2-M^2}{2\pi e^2}\sigma_{\gamma N}(w^2).
\eeqn

The integral over the electron's angles can be performed analytically, and 
we are left with the integral over the lab photon energy 
$\omega=\frac{w^2-M^2}{2M}$:
\beqn
{\rm Im}A_1^{inel}=\frac{1}{4\pi^2}
\frac{M}{E_{lab}}
&&\int_0^{E_{lab}} \!\!\!\!\!\!\!\! d\omega\omega\sigma_{\gamma N}(\omega)
\label{eq:ima1_inelast}\\
&&\times\ln\left[\frac{Q^2}{m^2}\left(\frac{E_{lab}}{\omega}-1\right)^2\right]
\nn,
\eeqn
with $E_{lab}=\frac{s-M^2}{2M}$ is the lab electron energy.
Photoabsorption cross section has been measured from threshold to high energies
for various nuclei, and it is known to approximately scale as the atomic number
of the nucleus. Therefore, the inelastic contribution to $A_n$ scales 
approximately as $\frac{A}{Z}$, while it scales as $Z$ for the elastic 
contribution. 

\section{Results and Discussion}
\label{results}
We now present our results for the analyzing power. 
We combine Eqs.(\ref{eq:ima1_elast}) with Eq.(\ref{eq:an_general}) for the 
contribution of the elastic intermediate state.
For the inelastic contribution, additional input is required. We have 
calculated the imaginary part of the amplitude $A_1$ 
by taking the exact forward limit for the nuclear Compton amplitude 
where the optical theorem is applicable. To depart from zero scattering 
angle, we have to make an assumption about the $t$-dependence of Compton 
scattering amplitude. In the case of the proton, Refs. 
\cite{afanasev1} and \cite{javvcs} use the slope of the differential 
Compton cross section for the proton target \cite{comptonslope} 
known for $-t=Q^2\leq1$GeV$^2$, 
\beqn
\frac{d\sigma}{dt}\approx\left[\frac{d\sigma}{dt}\right]_{t=0}\times e^{Bt}
\eeqn
with $B\approx 8$ GeV$^{-2}$.
Since the differential cross section is related to the amplitude squared, 
the $t$-dependence is naturally modelled by
\beqn
{\rm Im}A_1(\nu,Q^2)\approx{\rm Im}A_1(\nu)\times e^{-BQ^2/2}
\eeqn

Generalizing this approach to the case of the nuclear target, we can write for 
the analyzing power:
\beqn
A_n^{inelast}&\approx&-\frac{1}{4\pi^2}\frac{m_e}{E_{lab}}\frac{M}{\sqrt{s}}
\frac{A}{Z}\frac{g_N(Q^2)}{F_N(Q^2)}\tan\frac{\theta_{c.m.}}{2}\\
&\times&
\int_0^{E_{lab}} \!\!\!\!\!\!\!\! d\omega\omega\sigma_{\gamma p}(\omega)
\ln\left[\frac{Q^2}{m^2}\left(\frac{E_{lab}}{\omega}-1\right)^2\right]
\nn
\eeqn
where $g_N(Q^2)$ is the phenomenological Compton form factor for a 
nucleus $N$, and we made use of an approximate scaling of the photo absorption 
cross section with the atomic number $A$.
Unfortunately, the $t$-dependence of Compton data is not known for nuclei. 
Since the $t$-dependence of the elastic form factors of nuclei is much steeper 
than that of the nucleon, we also expect that the slope of Compton differential 
cross section should be much steeper, as well. 
Therefore, to provide an adequate 
prediction for the inelastic states contribution to the analyzing power at a 
non-zero scattering angle, we make a substitution in the above formula:
\beqn
\frac{g_N(Q^2)}{F_N(Q^2)}\rightarrow\frac{g_p(Q^2)}{F_1^p(Q^2)}
\eeqn
with $g_p(Q^2)=Exp[-\frac{B}{2}Q^2]$ and $F_1^p$ the proton Dirac form factor.

If we assume that the photoabsorption cross section is a constant in energy 
(which is roughly the case at energies above the resonance region, say, 
$\omega\geq2.5$ GeV, with $\sigma_{\gamma p}\approx0.1$ mbarn), the 
integration can be performed analytically, and we obtain a simple formula 
\beqn
A_n^{inelast}&\approx&
A_n^0\frac{g_N(Q^2)}{F_N(Q^2)}\tan\frac{\theta_{c.m.}}{2}
\left(\ln\frac{Q^2}{m^2}-2\right)
\eeqn
where $A_N^0=-\frac{m_e E_{lab}\sigma_{\gamma p}}{8\pi^2}\frac{M}{\sqrt{s}}
\frac{A}{Z}\approx -4$ ppm for lead. 
This result is analogous to that of Refs.\cite{afanasev1} and \cite{javvcs} 
obtained for the spin-$\frac{1}{2}$ target.
Analyzing this formula for a heavy nucleus, we can deduce the energy 
dependence at very forward angles, where the Compton slope is irrelevant:
\beqn
A_n\sim A_n^0 \tan\frac{\theta_{c.m.}}{2}
\left(\ln\frac{4E_{lab}^2}{m_e^2}-2+2\ln\sin\frac{\theta_{c.m.}}{2}\right)
\eeqn
\indent
Having in mind that $A_n^0$ defined above grows linearly with energy, 
we see that at fixed (forward) angle, the analyzing power behaves as $E\ln E$. 
At high energies, the phenomenological $t$-dependence tends to partially cancel 
this growth. On the other hand, for fixed momentum transfer, the analyzing 
power is practically independent on the beam energy, as was noticed in
\cite{afanasev1}.

\begin{figure}[h]
{\includegraphics[height=5.5cm]{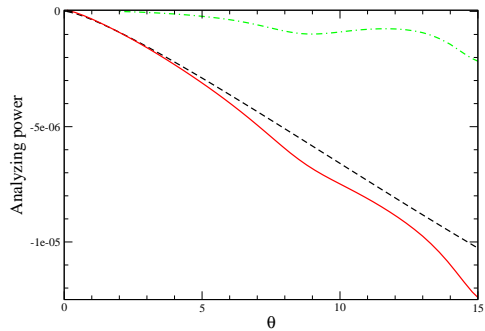}}
\caption{Analyzing power on $^{208}Pb$ at the electron beam energy of 850 MeV 
as function of the c.m. scattering angle in degrees.
Contributions from elastic (dash-dotted) and inelastic (dashed) intermediate 
states are shown, as well as their sum (solid).}
\label{fig:bn_lead}
\end{figure}

\begin{figure}[h]
{\includegraphics[height=5.5cm]{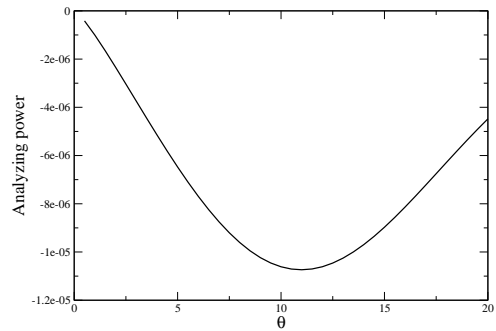}}
\caption{Inelastic contribution to analyzing power on $^{208}Pb$ at the 
electron beam energy of 2.7 GeV as function of the c.m. scattering angle in 
degrees.}
\label{fig:bn_lead2p7}
\end{figure}

We present the results for the $A_n$ for two different spin-0 nuclei. 
In Fig. \ref{fig:bn_lead}, we display the analyzing power on $^{208}Pb$ in the 
kinematics of the PREX experiment, 850 MeV beam and forward angles. 
It can be seen 
that the inelastic contributions give the main contribution, although the 
elastic contribution is also not negligible. The sum of the two leads to 
approximately $-4 ppm$ at 6 degrees. The elastic curve in 
Fig. \ref{fig:bn_lead} corresponds not to the calculation presented in this 
paper, but to the calculation of Ref. \cite{chuck} that sums the Coulomb 
distortion effects to all orders in $Z\alpha_{em}$. 
In Fig. \ref{fig:bn_lead2p7}, we display the inleastic contribution to $A_n$ on 
lead at forward angles and a higher energy of 2.7 GeV. The elastic contribution 
is not shown, as it is very small in those kinematics. 
For the $^4$He target, the elastic contribution is largely suppressed, both 
by a smaller nuclear charge than in the case of lead and by kinematics. 
The only sizeable contribution comes from the inelastic intermediate states 
and accounts for about $-10 ppm$ at 10 degrees c.m. scattering angle. 
This result closely agrees with the calculation of Ref. \cite{afanasev2}.
The exact number has to be taken with care, as it relies on a model-dependent 
$t$-slope that was taken the same as for the proton. This model should 
work at very small values of $Q^2$, but will fail at larger values. Whether or 
not the point $Q^2\approx 0.1$ GeV$^2$ is inside this reliable range, is 
definitely worth a future study. 
\begin{figure}[h]
{\includegraphics[height=5.5cm]{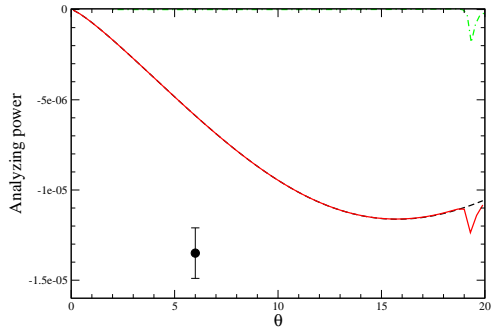}}
\caption{Analyzing power on $^{4}He$ at the electron beam energy of 3 GeV 
as function of the c.m. scattering angle in degrees.
Contributions from elastic (dash-dotted) and inelastic (dashed) intermediate 
states are shown along with the sum of the two (solid). The data point is from 
\cite{HAPPEX}.}
\label{fig:bn_he4}
\end{figure}
Finally, we discuss the quality of the leading order in $\alpha_{em}$ 
approximative result for the analyzing power of Eq.(\ref{eq:ima1_elast})
by comparing it to the full result of Ref.\cite{chuck}. 
This comparison is shown in Figs. \ref{fig:bn_he4_elast} and 
\ref{fig:bn_lead_elast} for $^4$He and $^{208}$Pb target, respectively. 
The expansion is performed in ``small'' parameter $Z\alpha_{em}$, thus 
it is expected to work well for helium, but not for lead where 
$Z\alpha_{em}\approx0.6$. 
Indeed, Figs. \ref{fig:bn_he4_elast} and \ref{fig:bn_he4_elast_zoom} 
demonstrate that for the whole interval in 
the scattering angle, the agreement between the two calculations is good, apart 
from the vicinity of the diffraction minimum in the $^4$He elastic form factor 
that enters the denominator of Eq.(\ref{eq:ima1_elast}). 
The leading order form factor is exactly zero in the diffraction minimum, 
while this minimum is partially filled by including Coulomb distortion effects 
in Ref.\cite{chuck}. 
\begin{figure}[h]
{\includegraphics[height=5.5cm]{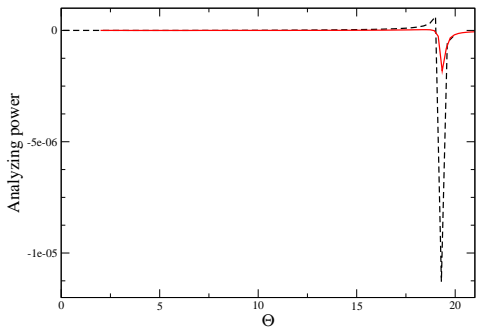}}
\caption{Elastic contribution to the analyzing power on $^{4}He$ at the 
electron beam energy of 3 GeV as function of the c.m. scattering angle in 
degrees. The leading order contribution (dashed curve) is compared to the full 
result (full curve) from Ref. \cite{chuck}.}
\label{fig:bn_he4_elast}
\end{figure}
\begin{figure}[h]
{\includegraphics[height=5.5cm]{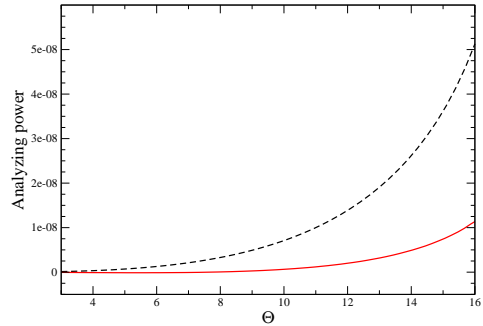}}
\caption{Zoomed version of Fig. \ref{fig:bn_he4_elast}}
\label{fig:bn_he4_elast_zoom}
\end{figure}
For lead, the agreement between the two calculations is unsatisfactory, and 
the elastic contribution to the analyzing power is relatively large, so 
it is necessary to include the higher orders, as well. 
\begin{figure}[h]
{\includegraphics[height=5.5cm]{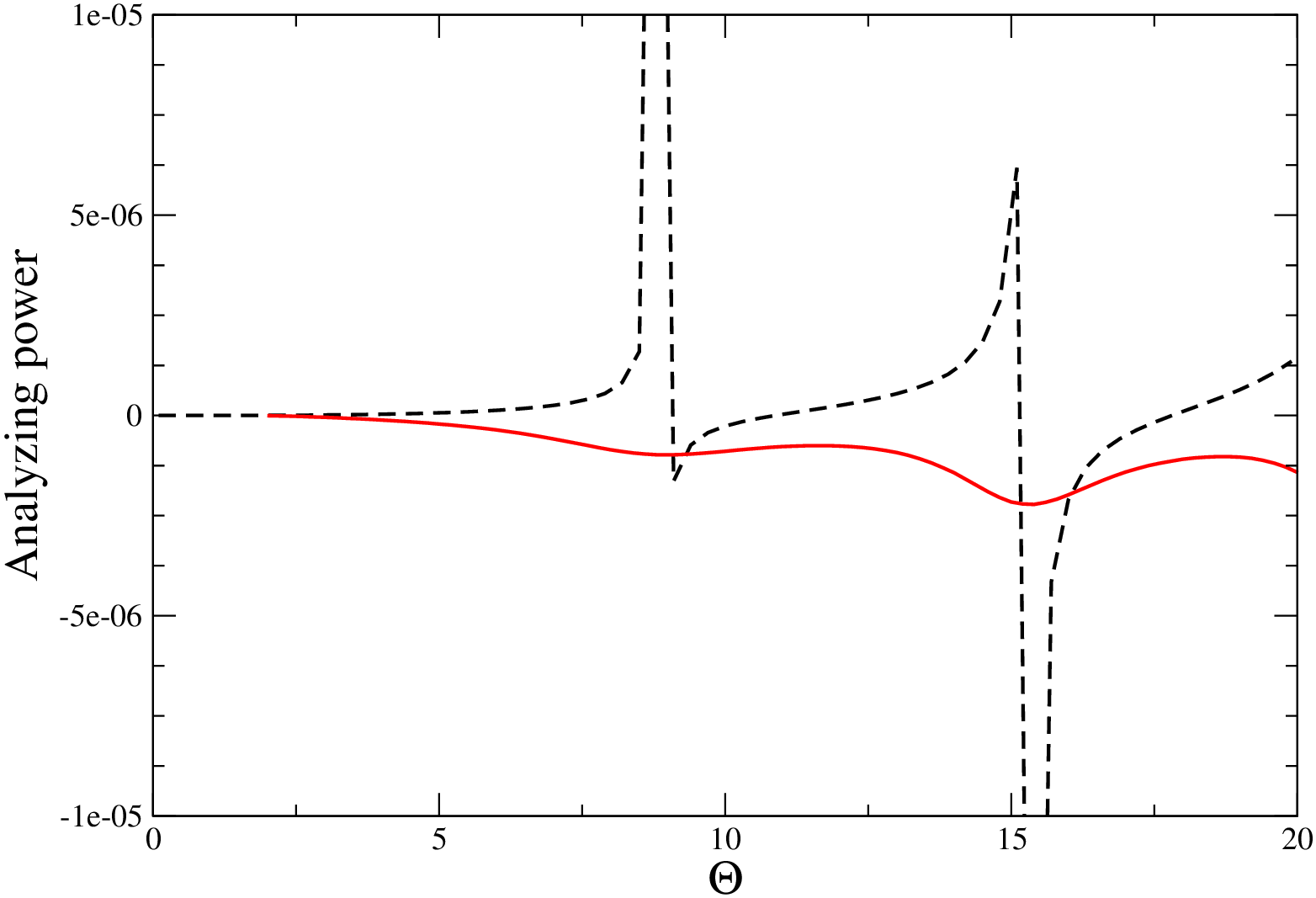}}
\caption{The same as in Fig. \ref{fig:bn_he4_elast} for the case of $^{208}Pb$.}
\label{fig:bn_lead_elast}
\end{figure}
We quote some of our numerical results in the kinematics of HAPPEX and PREX 
expreriments in Table \ref{tab}.

\begin{table}[ht]
\vspace{1cm}
   \begin{tabular}{|l|l|l|l|l|l|}
\hline
$\Theta_{c.m.}$(deg)  & \multicolumn{2}{|c|}{$A_n$ (ppm)} & $\Theta_{c.m.}$(deg)  & \multicolumn{2}{|c|}{$A_n$ (ppm)} \\
\hline
& $^4$He$\;\;\;\;\;\;$ & $^{208}$Pb & & $^4$He$\;\;\;\;\;\;$ & $^{208}$Pb \\
\hline
$0.5^\circ$ & & -0.09 & & & \\
$1.0^\circ$ & -0.72 & -0.33 & $11.0^\circ$ & -10.13 & -8.12 \\
$2.0^\circ$ & -1.68& -0.91 & $12.0^\circ$ & -10.68 & -8.85 \\
$3.0^\circ$ & -2.71 & -1.57 & $13.0^\circ$ & -11.11 & -9.75 \\
$4.0^\circ$ & -3.77 & -2.31 & $14.0^\circ$ & -11.41 & -10.98 \\
$5.0^\circ$ & -4.83 & -3.10 & $15.0^\circ$ & -11.58 & -12.43 \\
$6.0^\circ$ & -5.87 & -3.97 & $16.0^\circ$ & -11.61 & -12.94 \\
$7.0^\circ$ & -6.88 & -4.93 & $17.0^\circ$ & -11.50 & -13.05 \\
$8.0^\circ$ & -7.82 & -5.94 & $18.0^\circ$ & -11.28 & -13.39 \\
$9.0^\circ$ & -8.69 & -6.82 & $19.0^\circ$ & -11.06 & -13.98 \\
$10.0^\circ$ & -9.46 & -7.48 & $20.0^\circ$ & -10.73 & -14.97 \\
\hline
   \end{tabular}
\caption{Results for the analyzing power on $^4$He for 3 GeV beam energy and 
on $^{208}$Pb for 855 MeV beam energy in forward kinematics.}
\label{tab}
\end{table}

In summary, we considered elastic scattering of electrons off the spin-0 
nuclear target. The analyzing power for this scattering process is related 
to the imaginary part of the scattering amplitude, and thus requires an 
exchange of at least two photons. On one hand, the elastic intermediate state 
contribution is due to Coulomb distortion and can be calculated to all orders 
in the electromagnetic coupling constant \cite{chuck}. 
Another approach capitalizes on the fact  that the imaginary part of the 
forward Compton amplitude is related by the optical theorem to the total 
photo absorption cross section. Photoabsorption was measured on many nuclear 
targets, and we use it as input along with the $t$-dependence of the 
differential Compton cross section which is needed in order to depart from the 
exact forward limit. 
We applied this approach to $^4 He$ and $^{208}Pb$ nuclei in the kinematics 
of present parity-violation experiments and found that the analyzing power 
is negative in both cases and is about $-10ppm$ and $-4ppm$, respectively. 
The analyzing power is relatively large.  Experimentalists should take care to 
ensure that it does not contribute a large systematic error to the extraction 
of parity violating observables.

We showed that the account of Coulomb distortions to all orders in 
$Z\alpha_{em}$ modifies significantly the elastic contribution 
to the analyzing power for $^{208}Pb$. 
At the moment, only the leading order inelastic contribution was calculated, 
however it is plausible to assume that also this can be substantially 
modified, although not at the order-of-magnitude level, by the inclusion of the 
higher orders effects. This issue should be addressed in the future.

\acknowledgments
This work was supported in part by the US NSF grant PHY 0555232 (M.G.) 
and by DOE grant DE-FG02-87ER40365 (C.J.H.)

\end{document}